\documentclass[prl,final,twocolumn,showpacs,showkeys]{revtex4}

\usepackage{graphics}
\usepackage{graphicx}

\newcommand{\etal}{ et al. }

\newcommand{\be}{\begin{equation}}
\newcommand{\ee}[1]{\label{#1} \end{equation}}
\newcommand{\ba}{\begin{eqnarray}}
\newcommand{\ea}[1]{\label{#1} \end{eqnarray}}
\newcommand{\nl}{\nonumber \\}

\newcommand{\td}[2]{ \frac{d#1}{d#2} }
\newcommand{\pd}[2]{ \frac{\partial #1}{\partial #2} }

\newcommand{\infi}{\int_{-\infty}^{+\infty}\limits\!}

\newcommand{\photo}{ \frac{dN}{k_{\perp}dk_{\perp}d\eta \, d\psi}}

\renewcommand{\cosh}{{\rm ch}}
\renewcommand{\sinh}{{\rm sh}}
\renewcommand{\tanh}{{\rm th}}

\begin{document}


\title{ \bf Unruh gamma radiation at RHIC? }


\author{{ Tam\'as S. Bir\'o} }
	\affiliation{Theory Division,  MTA KFKI Res.\ Inst.\ for Particle and Nuclear Physics, Budapest, Hungary}
\author{{ Mikl\'os Gyulassy} }
	\affiliation{Dept.\ of Physics,  Columbia University, NY, USA}

\author{{Zsolt Schram} }
	\affiliation{Dept.\ of Theoretical Physics, University of Debrecen, Hungary}

\date{\today}

\vspace{12mm}

\begin{abstract}
Varying the proposition
that acceleration itself would simulate a thermal environment, we investigate the semiclassical
photon radiation as a possible telemetric thermometer of accelerated charges. Based on
the classical Jackson formula we obtain the equivalent photon intensity spectrum stemming from
a constantly accelerated charge and demonstrate its resemblances to a thermal distribution for high
transverse momenta. 
The inverse transverse slope {\em differs} from the famous Unruh temperature: it is larger by a factor of $\pi$.
We compare the resulting direct photon spectrum with experimental data for Au-Au collisions at RHIC
and speculate about further, analytically solvable acceleration histories.
\end{abstract}


%


\maketitle




\section{Thermal Looking Classical Radiation from Accelerating Source}

It has been a puzzle to understand why thermal models supposing global
equilibrium, and hydrodynamics assuming local equilibrium are so
successful in describing single particle observables in high energy
heavy ion and even in elementary particle reactions.
Neither is it easy to understand intuitively how 
 accelerating frames produce ''thermal vacua'', with quanta distributed as if they
were connected to a heat bath -- a proposition first formulated
by Unruh \cite{UNRUH} and soon analyzed in relation to black hole thermodynamics
by Bekenstein \cite{BH1} and Hawking \cite{BH2,BH3}. Spectra of photons from a black hole
was obtained by Page \cite{PAGE}.
These early analyses in the seventies were followed by numerous attempts to harvest
the merits of this idea in relation with possible ways to quantum gravity \cite{WaldBook}.
In the present letter we calculate measurable consequences of this idea
in photon spectra observed in high energy heavy ion collisions.
The observation that transverse momentum spectra of direct gammas,
measured perpendicular to the beam line in collider experiments, are well fitted by exponential
distributions \cite{PHENIX}, are suggestive of  ''thermal''  heat bath 
sources in high energy nuclear collisions.
Theoretical calculations assuming the applicability of thermodynamic
concept are phenomenologically successful in describing
heavy ion experimental data \cite{Greiner,Cleymans,Hirano,Baier,Aurenche,Thoma,Srivastava,Sreekanth,Renk}.
Statistical predictions based on the assumption of a thermal source for also $pp$ collisions
at the LHC have recently been made by Becattini \etal \cite{BecattiniEPJC,BecattiniJPG}.
Alternatively, the proposition that the thermal spectra in $p+p$  and in heavy ion $A+A$ collisions were due to accelerating
frame concepts akin to the Unruh temperature of field theoretic vacua was made by Satz and Kharzeev 
\cite{Kharzeev1,KHARZEEV,SatzKharzeev,Dima-in-Book}.
However, a dynamical calculation demonstrating 
how thermal looking spectra emerge from accelerating systems was not provided.

In this letter we demonstrate how the classical electromagnetic 
field of a constantly accelerating charge
develops an approximate   thermal  distribution in the transverse photon spectrum while also exhibiting a uniform rapidity dependence usually interpreted in terms of longitudinally boost invariant Bjorken hydrodynamics. 
This calculation is entirely
classical with the Planck constant entering solely by converting the intensity
distribution in terms of the  number distribution of  equivalent photons
as each carrying  the energy $\hbar\omega$. Beyond this we do not need 
to refer to field theory,
quantization procedure or the vacuum state at all. 
The trajectory of pointlike classical charges
and the radiated photon field are calculated in the framework of special relativity,
without any actual reference to a general frame and related Killing vectors. 

We consider a simple point charge undergoing constant acceleration. 
We calculate the classical bremsstrahlung distribution
using the text book\cite{JACKSON14} formula for the radiation field amplitude:
\be
 \vec{A} = K \int e^{i\varphi} \td{}{t} \left[ \frac{\vec{n}\times(\vec{n}\times\vec{\beta})}{1-\vec{n}\vec{\beta}} \right] \, dt
\ee{JACKSON_FORMULA}
with $K^2 = {e^2}/{8\pi c^2}$ and  retarded phase $\varphi = \omega \left( t - {\vec{n}\vec{r}}/{c} \right)$.
Here the trajectory of the moving charge is given by $\vec{r}(t)$ and 
 $\vec{\beta} = {\vec{v}}/{c} = d{\vec{r}}/{cdt}$.
The relativistic motion of the charge can also be described in terms of a proper
time $\tau$. This defines a Lorentz factor
 $\gamma = {1}/{\sqrt{1-\vec{\beta}^2}} = d{t}/{d\tau}$.
This way the derivative of the retarded phase with respect to the proper time
is given as
\be
 \td{\varphi}{\tau} = \omega \, \gamma (1-\vec{n}\vec{\beta}).
\ee{PHASE_DERIV}
Without restricting the generality the velocity vector can be taken to point into the
first spatial direction, $\vec{\beta} = (v/c,0,0)$,
with the magnitude $v/c=\tanh\xi$, $\xi$ being the rapidity variable
in the direction of the motion. 
The original Unruh formula arises by a Fourier analysis of the  phase observed in the particular
detecting direction of zero angle.

In our calculations we use the vector
\be
 \vec{u} = \frac{\vec{n}\times(\vec{n}\times\vec{\beta})}{1-\vec{n}\vec{\beta}}
 	= \frac{v \sin\theta}{c-v\cos\theta} \vec{e},
\ee{DIPO_VELO}
with $\vec{e}=\partial\vec{n}/\partial\theta$ unit vector orthogonal to $\vec{n}$.
The Doppler factor is given as
\be
 \gamma(1-\vec{\beta}\vec{n}) = \cosh\xi - \sinh\xi \cos\theta.
\ee{DOPP_FAC_REL}
Next we define  $ \tanh\eta = \cos\theta$.
Due to the extreme relativistic kinematics 
$\eta = \ln {\rm ctg} ({\theta}/{2})$ is the {\em rapidity} of the photon.
By the virtue of this, one arrives at the following Doppler factor:
\be
 \gamma(1-\vec{\beta}\vec{n}) = \frac{\cosh(\xi-\eta)}{\cosh\eta}.
\ee{DOPP_XI}
Since this is proportional to the derivative of the retarded phase with respect to the source's rapidity,
\be
 \td{\varphi}{\tau} = \frac{g}{c} \td{\varphi}{\xi} = \omega  \frac{\cosh(\xi-\eta)}{\cosh\eta},
\ee{DIFFEQ_DOPPLER}
by integration with constant acceleration, $g$, we arrive at 
\be
 \varphi = \frac{\omega c}{g} \frac{\sinh(\xi-\eta)}{\cosh\eta}.
\ee{PHASE_XI}
Considering that $\omega = ck_{\perp}/\sin\theta = ck_{\perp} \cosh\eta$ and $\ell=c^2/g$ 
is a length characteristic to the magnitude of acceleration, the above result can be written in a simpler form:
\be
 \varphi = \ell k_{\perp} \, \sinh(\xi-\eta).
\ee{SIMPLER_PHASE_XI}
The vector $\vec{u}$ can be expressed as
\be
 \vec{u} = \frac{\beta\sin\theta}{1-\beta\cos\theta} \, \vec{e} = 
   \frac{\sinh\xi}{\cosh(\xi-\eta)} \, \vec{e},
\ee{U_DIP_MAG}
its derivative is given by
\be
 \td{\vec{u}}{\xi} = \frac{\cosh\eta}{\cosh^2(\xi-\eta)} \, \vec{e}.
\ee{DERIV_OF_U_XI}
The Lorentz covariant radiation intensity distribution
is evaluated here in terms of transverse wavenumber and 
rapidity variables. The correspondence to the familiar covariant
notation is established by considering the following four-vector
definitions:
\be
 u = (\gamma, \gamma\vec{v}) = (\cosh\xi,\sinh\xi,0,0)
\ee{U_FOUR_DEF}
for the four-velocity of the radiating charge,
\be
 k = (\omega, \omega\vec{n}) = k_{\perp} (\cosh\eta,\sinh\eta,\cos\psi,\sin\psi)
\ee{K_FOUR_DEF}
for the four-momentum of the photon and
\ba
 \epsilon_1 &=& \frac{1}{\omega}\pd{k}{\theta} = (0,-\sin\theta,\cos\theta\cos\psi,\cos\theta\sin\psi),
\nl
 \epsilon_2 &=& \frac{1}{\omega\sin\theta}\pd{k}{\psi} = (0,0,-\sin\psi,\cos\psi),
\ea{POLARIZATION}
for the polarization vectors. Using these covariant notations the familiar difference for
the amplitude is replaced by a continuous integral over rapidity
\be
 \left. \frac{\epsilon\cdot u}{k \cdot u} \right|_1^2 \longrightarrow 
 \int_1^2 \td{}{\xi} \left( \frac{\epsilon \cdot u}{k \cdot u} \right) \, \, d\xi.
\ee{REPLACEMENT}
The Jackson formula (\ref{JACKSON_FORMULA}) is more in taking into account a continuously changing phase along
the path of the moving charge, arriving at
\be
 {\cal A} = K \int e^{i\varphi} \td{}{\xi} \left( \frac{\epsilon \cdot u}{k \cdot u} \right) \, \, d\xi. 
\ee{COVAR_AMPLITUDE}
Finally the spectrum is given as
\be
 \photo = 2 \, \left| {\cal A} \right|^2.
\ee{SPECT_COVAR}
In any case the description of the linear motion of the source of the radiation is given by
the proper acceleration, $d\xi/d\tau = g(\xi)$. Accordingly the phase has to be obtained 
from the integration of the relation
\be
 \td{\varphi}{\tau} = \td{}{\tau} k\cdot x = k \cdot u = k_{\perp} \cosh(\xi-\eta).
\ee{COVAR_DIFF_PHASE}

The general trajectory with constant relativistic acceleration, the Unruh trajectory, is given
parametrically as
\ba
 t = t_0 + \frac{c}{g} \left( \sinh(\xi+\xi_0) - \sinh\xi_0 \right), 
\nl
 x_3 =  x_{03} + \frac{c^2}{g} \left( \cosh(\xi+\xi_0) - \cosh\xi_0 \right).
\ea{UNRUH_TRAJECT}
with $ \beta = {v}/{c} = \tanh(\xi+\xi_0)$ and $ \xi = {g}\tau/c$.
The phase space factor of the radiation can be expressed by
using the $k_{\perp}=\omega\sin\theta/c$ transverse momentum and the rapidity $\eta$:
One obtains $d\omega d\Omega=c(dk_{\perp}d\eta)/\cosh\eta$. This leads to the following
rewriting of the differential spectrum for the photon number: 
\be
  \photo = \frac{2}{\hbar k_{\perp}^2\cosh^2\eta} \left|\vec{A} \right|^2.
\ee{PHOTON_SPECTRUM}
By determining the amplitude of the radiation it is noteworthy that the
derivative of the vector $\vec{u}$ with respect to time and the integration
with respect to the same time can be replaced by any variable. 
The amplitude is given by using the above expressions for $\varphi$ and $\vec{u}$ as functions of $\xi$:
\be
 \vec{A} = K\vec{e} \infi e^{i \ell k_{\perp} \sinh(\xi-\eta)} 
	\frac{\cosh\eta}{\cosh^2(\xi-\eta)} \, d\xi.
\ee{VECA_OF_INT_XI}
From this one obtains the following analytic result
\be
 \vec{A} = 2 K\vec{e} \:  \frac{\omega c}{g}  \: \: K_1\left(\ell k_{\perp}  \right)
\ee{VECA_ANAL}
with the Bessel K-function.
The spectrum of equivalent photons is given by
\be
 \photo = \frac{4 \alpha_{EM}}{\pi} \, \ell^2 \, K_1^2(\ell k_{\perp}),
\ee{SPECTRUM}
with $\alpha_{EM}=e^2/4\pi\hbar c$.

For heavy ion collisions basically two limiting approximations can be considered: 
the coherent square of sum of amplitudes and the incoherent sum over squared elementary amplitudes. 
For $k_T> 1$ GeV and $R\sim 7$ fm  nuclear radii,
coherence is negligible and the incoherent sum leads to an effective charge factor
\be
 F = 2 Z(b)
\ee{COH_FACTOR}
with $Z(b)$ being the proton participant number of each nucleus
colliding at impact parameter $b$. 
$Z(0)=79$ most central $b=0$ for Au beams, while $\langle Z \rangle \approx 55$ for the $0-20\%$ centrality trigger bin
considered in the PHENIX experiment.


\section{Thermal interpretation: not exactly Unruh}

For low $k_{\perp}$ the spectrum (\ref{SPECTRUM}) 
resembles the {\em conformal} $1/k_{\perp}^2$ limit independent of
$\ell$. For high $k_{\perp}$ on the other hand it develops an approximate 
exponential tail
resembling {\em thermal} fireball models:
\be
 \photo  \rightarrow  F \, 2\alpha_{EM} \, 
	\frac{\ell}{k_{\perp}} \,  e^{-2\ell k_{\perp}}.
\ee{HIGH_KT_SPEC}
The equivalent temperature is seen from the exponent 
and can be interpreted if we wish as $\pi$ times the Unruh-temperature, $T_U$:
\be
 T = \frac{\hbar c}{2k_B \ell} = \frac{\hbar}{2k_B c} g = \pi \: T_U.
\ee{TEMPER_HIGH_KT}
Note however that this photon spectrum is uniform in rapidity
resembling  Bjorken hydrodynamic scenario.
The uniform rapidity distribution is nevertheless truncated 
at large $\eta$  if we consider a  finite range of time for which $g$ is constant.

Although the photon spectrum obtained above does contain an exponential tail,
and defines an {\em effective}  transverse temperature, 
it is clear from this specific dynamical model that no thermally equilibrated
source was ever involved. It only ``appears as if'' a locally
equilibrated Bjorken hydrodynamic fluid  sourced the photons.
While the classical radiation spectrum of photons
from a constantly accelerated point charge shows an exponential tail, and by
the virtue of this defines a temperature related to the acceleration, $g$, 
the details of the resulting photon rapidity and $k_{\perp}$ distribution
are {\em not} identical to that assumed with thermal models.

Instead of a thermal Bessel K1 function, its square occurs, and instead of the Unruh temperature
its $\pi$ times enlarged value governs the tail. 
It is therefore important
to understand better the classical EM result and the role of the K-Bessel 
function. The latter emerges not only from the EM Fourier integral above,
but also from the relativistic statistical Gibbs distribution.
Also the angle- and rapidity variables,
and integrals over them, occur in relativistic calculations 
with an assumed collective fluid velocity field.
It is possible to reinterpret
eq.(22) in terms of a J\"uttner-type representation by using the identity
\be
  	\infi\!d\zeta \, K_2\left( z \, \cosh\zeta \, \right)
	 = K_1^2\left( \frac{z}{2} \right).
\ee{K2_K1_SQUARED}
Utilizing this relation we get
\be
 \photo = F \frac{4\alpha_{EM} \ell^2}{\pi} \, 
  	\infi\!d\zeta \, K_2\left( 2\ell k_{\perp}\cosh(\zeta-\eta) \, \right).
\ee{PHOTO_AS_FLOW_THERMAL}
This shows that ordinary bremsstrahlung for a uniform accelerating
charge can indeed be thought of a superposition of 
fluid cells distributed with uniformly distributed  collective fluid rapidity, 
referred to longitudinal flow, evaluated at a fix freeze-out proper time
when the local temperature has cooled to eq.(25). It is rather remarkable
how two completely different dynamical frameworks, classical EM and classical hydrodynamics, 
can lead to closely identical ``predictions'' or more properly identical
fits to spectra resembling eqs.(22,27).

Finally we consider a finite period of uniform acceleration.In that
case  the rapidity integration is done between finite limits:
\be
 \photo = F \frac{\alpha_{EM}}{\pi} \, \frac{1}{k_{\perp}^2} \,
	\left| \int_{w_1}^{w_2}\limits\! e^{i\ell k_{\perp} w} \frac{dw}{(1+w^2)^{3/2}} \right|^2,
\ee{FINITE_PHOTO_SPEC}
with $w_{1,2}=\sinh(\xi_{1,2}+\xi_0-\eta)$. 
Numerically a rapidity interval from $-3$ to $+3$ suffices to approach
the idealized $|w_i|\rightarrow \infty$ limit.
Thus at high energies at RHIC and LHC where the total rapidity interval exceeds 10,
the end effects can be safely neglected at central rapidities $|\eta| \le \, \sim 1$

In the low energy opposite limit, on the other hand, the integral can be
estimated for $w_2-w_1$ small, and the
integrand evaluated at $w_0=(w_1+w_2)/2$ to get
\be
  \photo \approx F \frac{\alpha_{EM}}{\pi} \, \frac{1}{k_{\perp}^2} \,
	\frac{1}{\cosh^4(\chi)}	
\ee{LOW_KT_SHORT_TIME}
with $\chi=(\xi_1+\xi_2)/2+\xi_0-\eta$. For a symmetric scenario of decelerating trajectory
$\chi= \eta$ and the above formula resembles a Landau hydrodynamic 
scenario in terms of the photon energy:
\be
 \omega \frac{d^3N}{d^3k} = \frac{d^3N}{\omega d\omega d\Omega} \approx  F \frac{\alpha_{EM}}{\pi} \,
	\frac{1}{\omega^2 \cosh^2\eta}.
\ee{LANDAU_SPECTRUM}

\section{Comparing various acceleration profiles}

For comparison we consider several simple analytically solvable trajectories.
One engineered to give a pure exponential is defined by
$g=g_0\cosh\xi(1+\cosh\xi)$ and $t=\tanh(\xi/2)$ while $v/c=\tanh(\xi)$. It leads to a 
pure exponential spectrum at $\theta=90$ degree ($\eta=0$),
\be
 \left. \photo \right|_{\eta=0} \, = \, F \frac{\alpha_{EM}}{\pi} \, \ell^2 e^{-2\ell k_{\perp}}.
\ee{2D_SPECTRUM}
This has, however, an unrealistic acceleration profile, 
having its minimum at $\xi=0$ as $g_{{\rm min}}=2g_0$
while $\ell=c^2/g_0$.

A more realistic example is 
implicitly given by the trajectory $t={\rm tg}(\pi v/2)$ and
$x=1+v\,{\rm tg}(\pi v/2) + (2/\pi) \ln \cos(\pi v/2)$. In this case the acceleration is
mostly zero, it emerges up to $g_0\pi/2$ at $\xi=0$. The resulting photon spectrum
is 
\be
 \photo = F \frac{\alpha_{EM}}{\pi} \, \frac{1}{k_{\perp}^2}  e^{-2\ell k_{\perp}}.
\ee{1D_SPECTRUM}
Finally we note that there is always a thermal looking part of the photon spectrum
beyond the transverse momentum for which the retarded phase is larger than $\pi$.
This is achieved for
\be
 k_{\perp}^* = \frac{\pi}{c\Delta\tau} \frac{\Delta\xi}{\sinh\Delta\xi}
\ee{PHASE_PI_KPERP}
if estimating the deceleration by $g/c \approx \Delta\xi/\Delta\tau$ for short stopping.
It is noteworthy that even for $\Delta\xi\rightarrow 0$ there is a thermal tail, for
finite rapidity acceleration histories this region is pushed towards lower $k_{\perp}$ values.

\begin{figure}
\begin{center}
        \includegraphics[width=0.30\textwidth,angle=-90]{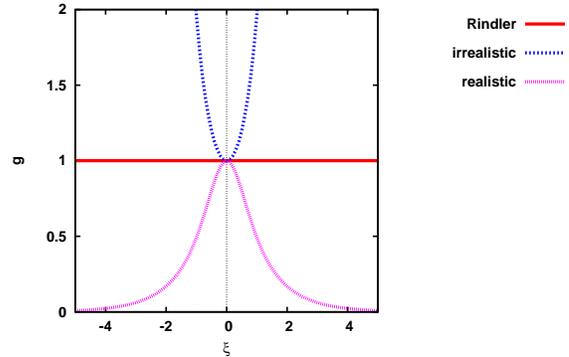}
\end{center}

\caption{ \label{FIG:ACCEL} (Color online)
  Acceleration profiles against the rapidity of a point charge.
  The photon spectrum is analytically calculable for these cases.
}
\end{figure}
\begin{figure}
\begin{center}
        \includegraphics[width=0.34\textwidth,angle=-90]{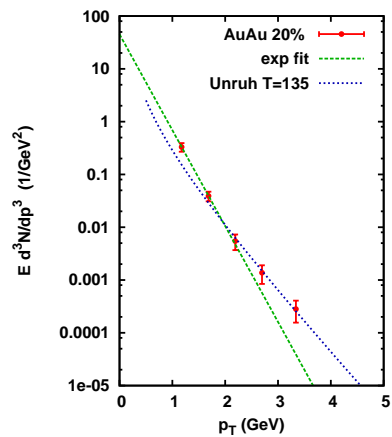}
\end{center}

\caption{ \label{FIG:PHENIX} (Color online)
  Comparison of the Maxwell/Unruh/Bjorken like
 bremsstrahlung formula eq.(22) with RHIC PHENIX data on direct photons
  in 20\% central Au-Au collisions for the constant $g=2\pi T_U$ scenario applying $T_U=135$ MeV
  and the effective charge $Z(b)=55$ (dotted line). 
  The pure exponential fit by the PHENIX group fitting the temperature to $T=240$ MeV
  and the amplitude to $A=45.2$ is indicated by the dashed line.
}
\end{figure}

In Figure \ref{FIG:ACCEL} we present the proper acceleration as a function of the 
rapidity of the source charge for three different, analytically solvable scenarios
discussed above. The realistic one starts and ends with low acceleration magnitude
and reaches its maximum at $\xi=0$ simulating the strongest medium effect
in the laboratory frame. 

In Figure \ref{FIG:PHENIX} we compare transverse gamma spectra arising from
the simplest (constant acceleration) Unruh like scenario bremsstrahlung
with Au-Au experimental data at RHIC \cite{PHENIX}. According to the publication
cited above at low $p_T=\hbar k_T$ the spectrum is dominated by a ''thermal'' contribution,
fitted by a pure exponential form $A\exp(-p_T/T)$ with $A=45.2$ and $T=240$ MeV. 
For comparison we show the Maxwell/Unruh/Bjorken 
result with the corresponding factor 
$\ell=1/(2\pi T_U)$ in the proper units, using the temperature of $T_U=135$ MeV. 
Here the magnitude is given once the deceleration factor $g$ -- or the corresponding Unruh
temperature, $T_U$ -- is fixed. Our primary aim here is not to fit in detail the 
experimental data but simply to point out the degeneracy of various theoretical models 
in terms of bremsstrahlung, Unruh , and Bjorken 1+1D perfect fluid 
hydrodynamics interpretations.
The open question is the uniqueness of the interpretation of
direct photon data in terms of purely thermal concepts.
How much of the observed spectra is due to accelerating classical field effects?
Here we limited the consideration to pure classical EM observables.
In the Color Glass Models\cite{McLerran:2010ub} large classical chromo field
effects were also shown to be able to fit certain
hadronic observables in heavy ion reactions similar to thermal/hydrodynamic models.
Only through detailed systematic studies of $\sqrt{s}, A,$ and impact parameter $b$ 
scaling in $pp, pA, AA$ of a wide variety of experimentally observable 
data can these, in their physical background differing, 
pictures and mechanisms be properly sorted out.


\vspace{3mm}
\section*{Acknowledgment}
This work has been supported by the Hungarian National Research Fund,
OTKA (K68108), by a Hungarian-South-African TeT project (TeT-10-1-2011-0061, ZA-15/2009),
and by the T\'AMOP 4.2.1./B-09/1/KONV-2010-0007 project,
co-financed by the European Union and European Social Fund. 
M.G. acknowledges also support from Division of Nuclear Science, 
U.S. Department of Energy, under Contract DE-FG02-93ER-40764. 
T.S.B. and Z.S. thank the  support by the Helmholtz International  
Center for FAIR within the framework of the LOEWE program 
(LandesOffensive zur Entwicklung Wissenschaftlich-\"okonomischer  
Exzellenz) launched by the State of Hesse.
Discussions with Prof.~I.~R\'acz are gratefully acknowledged.




\begin{thebibliography}{xxx}


\bibitem{UNRUH}
{W.~G.~Unruh},
{Phys. Rev. D}{\bf 14}{(1976)}{870}.

\bibitem{BH1}
{J.~D.~Bekenstein},
{Phys. Rev. D}{\bf 7}{(1973)}{2333},

\bibitem{BH2}
{J.~M.~Bardeen, B.~Carter and S.~W.~Hawking},
{Comm. Math. Phys.}{\bf 31}{(1973)}{161}.

\bibitem{BH3}
{S.~W.~Hawking},
{Comm. Math. Phys.}{\bf 43}{(1975)}{199}.

\bibitem{PAGE}
{D.~N.~Page},
{Phys. Rev. D}{\bf 13}{(1976)}{198}.



\bibitem{WaldBook}      
R.~M.~Wald:
Quantum Field Theory in Curved Spacetime and Black Hole Thermodynamics,
University of Chicago Press. 1994


\bibitem{PHENIX}
{A.~Adare  \etal for PHENIX},
{Phys. Rev. Lett}{\bf 104}{(2010)}{132301}


\bibitem{Greiner}
{A.~Dumitru, U.~Katscher, J.~A.~Maruhn, H.~Stocker and W.~Greiner},
{Phys. Rev. C}{\bf 51}{(1995)}{2166}

\bibitem{Cleymans}
{J.~Cleymans, K.~Redlich and D.~K.~Srivastava},
{Phys. Rev. C}{\bf 55}{(1997)}{1431}

\bibitem{Hirano}
{T.~Hirano, S.~Muroya and M.~Namiki},
{Prog. Theor. Phys.}{\bf 98}{(1997)}{129}

\bibitem{Baier}
{R.~Baier, M.~Dirks, K.~Redlich and D.~Schiff},
{Phys. Rev. D}{\bf 56}{(1997)}{2548}

\bibitem{Aurenche}
{P.~Aurenche, F.~Gelis, R.~Kobes and H.~Zakaret},
{Phys. Rev. D}{\bf 58}{(1998)}{085003}

\bibitem{Thoma}
{F.~D.~Steffen and M.~H.~Thoma},
{Phys. Lett. B}{\bf 510}{(2001)}{98},
{Erratum-ibid.}{\bf 660}{(2008)}{604}

\bibitem{Srivastava}
{R.~Chatterjee, D.~K.~Srivastava and U.~Heinz},
{Proc. of Eilat 2008, Particles and nuclei (PANIC08), 534}

\bibitem{Sreekanth}
{J.~R.~Bhatt, H.~Mishra and V.~Sreekanth},
arxiv: 1005.2756, 2010

\bibitem{Renk}
{R.~Chatterjee, H.~Holopainen, T.~Renk and K.~J.~Eskola},
{Phys. Rev. C}{\bf 83}{(2011)}{054908}


\bibitem{BecattiniEPJC}
{F.~Becattini, P.~Casterina, A.~Milov and H.~Satz},
{Eur. Phys. J. C}{\bf 66}{(2010)}{377}

\bibitem{BecattiniJPG}
{F.~Becattini, P.~Casterina, A.~Milov and H.~Satz},
{J. Phys. G}{\bf 38}{(2011)}{025002}


\bibitem{Kharzeev1}      
{D.~Kharzeev and H.~Satz}, 
{Phys. Lett. B}{\bf 334}{(1994)}{155}

\bibitem{KHARZEEV}      
{D.~Kharzeev and K.~Tuchin},
{Nucl. Phys. A}{\bf 753}{(2005)}{316}.

\bibitem{SatzKharzeev}  
{P.~Casterina, D.~Kharzeev and H.~Satz},
{Eur. Phys. J.  C}{\bf 52}{(2007)}{187}.

\bibitem{Dima-in-Book}	
{T.~Brasovenau, D.~Kharzeev and M.~Martinez}:
{\em In Search of QCD -- Gravity Correspondence},
{In ''The Physics of the Quark Gluon Plasma''},
{Eds. S.~Sarkar, H.~Satz, B.~Sinha}, {Springer}, {2009}.


\bibitem{JACKSON14}  
{D.~Jackson}: 
{\em Classical Electrodynamics},
Chap.~14,
Wiley, New York, 1975.


\bibitem{ACTA-DEB}	
{T.~S.~Bir\'o and Z.~Schram},
{Acta Physica Debrecina}{\bf XLV }{(2011)}{176}.




\bibitem{PHENIX_PRC}
{A.~Adare \etal},
{Phys. Rev. C}{\bf 81}{(2010)}{034911}. 






\bibitem{McLerran:2010ub}
  L.~McLerran,
 ~Acta~Phys.\ Polon.\  B {\bf 41}, 2799 (2010).
  F.~Gelis, E.~Iancu, J.~Jalilian-Marian, R.~Venugopalan,
  Ann.\ Rev.\ Nucl.\ Part.\ Sci.\  {\bf 60}, 463 (2010).



\end{thebibliography}
\end{document}